# Numerical investigation of a solar greenhouse tunnel drier for drying of copra

S. Sadodin [a], T. T. Kashani [a,*]

[a] *Department of Energy Conversion, Faculty of Mechanical Engineering, Semnan University, Semnan, Iran*

**Abstract**

A numerical investigation of a solar greenhouse tunnel drier (SGTD) has been performed. In the present study, the geometry of the tunnel roof is assumed semi-circular which is covered with a UV (200μ) stabilized polyethylene film. The simulated SGTD reduces moisture of copra from 52.2% to 8% in 55 h under full load conditions. A system of partial differential equations describing heat and moisture transfer during drying copra in the solar greenhouse dryer was developed and this system of non-linear partial differential equations was solved numerically using the finite difference method (FDM). The numerical solution was programmed in Compaq Visual FORTRAN version 6.5. The simulated results reasonably agreed with the experimental data for solar drying copra. This model can be used to provide the design data and is also essential for optimal design of the dryer. For instance the user is able to change the radiation properties of the roof cover for different materials of roof cover.

* Corresponding author. Tel.: +98 9352177236; fax: +98 2313325626.
E-mail address: ta_ka79@stu-mail.um.ac.ir

**Nomenclature**

| | | | |
|---|---|---|---|
| $A_c$ | area of the cover (m$^2$) | $M_p$ | Moisture content of product in the dryer model (db, decimal) |
| $A_f$ | area of the concrete floor (m$^2$) | $m_a$ | mass of air inside the dryer (kg) |
| $A_p$ | area of the product (m$^2$) | $m_c$ | mass of the cover (kg) |
| $a_w$ | water activity (–) | $m_p$ | mass of dry product (kg) |
| $C_{pa}$ | specific heat of air (J kg$^{-1}$ K$^{-1}$) | Nu | Nusselt number (–) |
| $C_{pc}$ | specific heat of cover material (J kg$^{-1}$ K$^{-1}$) | Re | Reynolds number (–) |
| $C_{pp}$ | specific heat of product (J kg$^{-1}$ K$^{-1}$) | rh | relative humidity (%) |
| $C_{pl}$ | specific heat of liquid water (J kg$^{-1}$ K$^{-1}$) | T | temperature of air in thin layer and equilibrium moisture models (°C) |
| $C_{pv}$ | specific heat of water vapour (J kg$^{-1}$ K$^{-1}$) | $T_a$ | drying air temperature in the dryer model (K) |
| D | average distance between the floor and the cover (m) | $T_{am}$ | ambient temperature (K) |
| $D_h$ | hydraulic diameter (m) | $T_c$ | cover temperature (K) |
| $D_p$ | thickness of the product (m) | $T_1$ | temperature at the last limit of the heat conduction inside the soil (K) |
| $F_p$ | fraction of solar radiation falling on the product (–) | $T_f$ | floor temperature (K) |
| H | humidity ratio of air inside the dryer (–) | $T_{in}$ | temperature of the inlet air of the dryer (K) |
| $H_{in}$ | humidity ratio of air entering the dryer (–) | $T_{out}$ | temperature of the outlet air of the dryer (K) |
| $H_{out}$ | humidity ratio of the air leaving the dryer (–) | $T_p$ | temperature of product (K) |
| h | parameter in the colour measurement (–) | $T_s$ | sky temperature (K) |
| $h_{c,c-a}$ | convective heat transfer between the cover and the air (Wm$^{-2}$ K$^{-1}$) | t | time (s) |
| $h_{c,f-a}$ | convective heat transfer between the floor cover and the air (Wm$^{-2}$ K$^{-1}$) | $U_c$ | overall heat loss coefficient from the cover to ambient air (Wm$^{-2}$ K$^{-1}$) |
| $h_{c,p-a}$ | convective heat transfer between the product and the air (Wm$^{-2}$ K$^{-1}$) | V | volume of the drying chamber (m$^3$) |
| $h_{r,c-s}$ | radiative heat transfer between the cover and the sky (Wm$^{-2}$ K$^{-1}$) | $V_a$ | air speed in the dryer (m s$^{-1}$) |
| $h_{r,p-c}$ | radiative heat transfer between the product and the cover (Wm$^{-2}$ K$^{-1}$) | $V_{in}$ | inlet air flow rate (m3 s$^{-1}$) |
| $h_w$ | convective heat transfer between the cover and the ambient (Wm$^{-2}$ K$^{-1}$) | $V_{out}$ | outlet air flow rate (m$^3$ s$^{-1}$) |
| $h_{d,f-g}$ | conductive heat transfer between the floor and the underground (Wm$^{-2}$ K$^{-1}$) | $V_w$ | wind speed (m s$^{-1}$) |
| $I_t$ | incident solar radiation (Wm$^{-2}$) | W | width of the dryer floor (m) |
| $k_a$ | thermal conductivity of air (Wm$^{-2}$ K$^{-1}$) | x | distance (m) |
| $k_f$ | thermal conductivity of the floor material (Wm$^{-1}$ K$^{-1}$) | $y_{exp,i}$ | experimental values |
| $k_c$ | thermal conductivity of the cover material (Wm$^{-1}$ K$^{-1}$) | $y_{pre,i}$ | predicted values |
| $L_p$ | latent heat of vaporization of moisture from product (J kg$^{-1}$) | $\alpha_c$ | absorptance of the cover material (–) |
| M | moisture content (%, db) | $\alpha_f$ | absorptance of the floor (–) |
| $M_0$ | initial moisture content of product (%, db) | $\alpha_p$ | absorptance of the product (–) |
| $M_e$ | equilibrium moisture content of product (%, db) | $\delta_c$ | thickness of the cover (m) |
| | | $\sigma$ | Stefan–Boltzmann's constant (W m$^{-2}$ K$^{-4}$) |
| | | $\rho_a$ | density of air (kg m$^{-3}$) |
| | | $\rho_c$ | density of the cover material (kg m$^{-3}$) |
| | | $\rho_p$ | density of the dry product (kg m$^{-3}$) |
| | | $\tau_c$ | transmittance of the cover material (–) |
| | | $\varepsilon_c$ | emissivity of the cover material (–) |
| | | $\varepsilon_p$ | emissivity of the product (–) |
| | | $\upsilon_a$ | kinematic viscosity of air (m$^2$ s$^{-1}$) |

# 1. Introduction

Copra is the richest source of oil (70%). Moisture content 52% in fresh coconuts is required to be reduced to 8% by drying to concentrate oil content. Copra is one of the most economically important fruit crops in the southern Iran. It is widely processed as a dried fruit for export to China and Taiwan. Copra is available year round as fresh and dried fruit. Copra is dried not only for preservation purposes, but also for modification of the taste, flavour and texture to meet consumer preferences and to increase market value of the product.

Traditional methods, mainly open-air drying of Copra, are commonly practiced in southern Iran. Drying of Copra in the traditional method of sun drying is susceptible to contamination by insects, dust and rain resulting in poor quality dried products. Also solar energy is utilized most inefficiently in traditional method of sun drying. In contrast, solar greenhouse tunnel driers have none of these disadvantages. The products of this kind of dryers have good quality.

Furthermore, solar drying is a renewable and environmentally friendly technology. Solar drying can be considered as an advancement of natural sun drying and it is a more efficient technique of utilizing solar energy (Bala, 1998; Muhlbauer, 1986). Many studies on natural convection solar drying of agricultural products have been reported (Exell and Kornsakoo, 1978; Zaman and Bala, 1989; Sharma et al., 1995; Oosthuizen, 1995). Considerable studies on simulation of natural convection solar drying of agricultural products and optimization have also been reported (Bala and Woods, 1994; Simate, 2003; Forson et al., 2007). However, the success achieved by natural convection solar dryers has been limited due to low buoyancy induced air flow.

Nevertheless, solar drying systems must be properly designed in order to meet particular drying requirements of specific products and to give optimal performance. Designers should investigate the basic parameters such as dimensions, temperature, relative humidity, airflow rate and the characteristics of products to be dried. However, full-scale experiments for different products, drying seasons and system configurations are sometimes costly and impractical. The development of a simulation model is a valuable tool for predicting the performance of solar drying systems.

Again, simulation of solar drying is essential to optimize the dimensions of solar drying systems and optimization techniques can be then used for optimal design of solar drying systems (Bala, 1998). A number of studies have been reported on greenhouse crop drying (Condori and Saravia, 1998; Garg and Kumar, 2000; Condori et al., 2001). Limited studies have been reported on modeling of a solar greenhouse dryer (Jain and Tiwari, 2004; Jain, 2005; Farhat et al., 2004). Kumar and Tiwari (2006) reported thermal modeling of jaggery drying in a natural convection solar greenhouse dryer. The model was programmed in MATLAB and predicted values and observed data agreed well.

# 2. Numerical procedure

*2.1. Mathematical modeling:*

A mathematical model was developed for predicting the performance of this type of dryer. The assumptions in developing the mathematical model are as follows:

(i) There is no stratification of the air inside the dryer.
(ii) Drying computation is based on a thin layer drying model.
(iii) Specific heat of air, cover and product are constant.
(iv) Absorptivity of air is negligible.
(v) Fraction of solar radiation lost through the north wall is negligible.

Schematic diagram of energy transfers inside the solar greenhouse dryer is shown in Fig.1 and the following heat and mass balances are formulated.

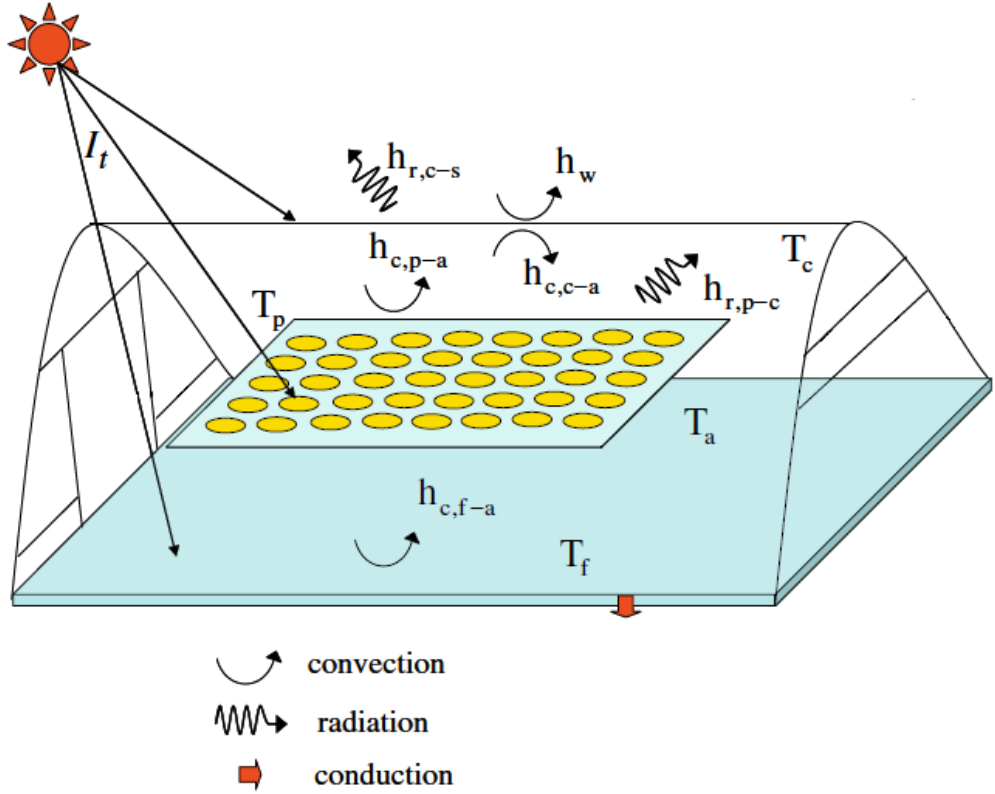

Fig. 1. Schematic diagram of energy transfers inside the solar greenhouse dryer

*2.1.1. Energy balance of the cover*

The balance of energy on the cover is considered as follows: Rate of accumulation of thermal energy in the cover = Rate of thermal energy transfer between the air inside the dryer and the cover due to convection + Rate of thermal energy transfer between the sky and the cover due to radiation + Rate of thermal energy transfer between the cover and ambient air due to convection + Rate of thermal energy transfer between the product and the cover due to radiation + Rate of solar radiation absorbed by the cover.

The energy balance of the polyethylene cover gives:

$$m_c C_{pc} \frac{dT_c}{dt} = A_c h_{c,c-a}(T_a - T_c) + A_c h_{r,c-s}(T_s - T_c) \\ + A_c h_w(T_{am} - T_c) + A_p h_{r,p-c}(T_p - T_c) + A_c \alpha_c I_t.$$

*2.1.2. Energy balance of the air inside the dryer*

This energy balance can be written as: Rate of accumulation of thermal energy in the air inside the dryer = Rate of thermal energy transfer between the product and the air due to convection + Rate of thermal energy transfer between the floor and the air due to convection + Rate of thermal energy gain of the air from the product due to sensible heat transfer from the product to the air + Rate of thermal energy change in the air chamber due to inflow and outflow of the air in the chamber + Rate of overall heat loss from the air in the dryer to the ambient air + Rate of solar energy accumulation inside dryer from solar radiation.

The energy balance in the air inside the greenhouse chamber gives:

$$m_a C_{pa} \frac{dT_a}{dt} = A_p h_{c,p-a}(T_p - T_a) + A_f h_{c,f-a}(T_f - T_a) \\ + A_p D_p C_{pv} \rho_p (T_p - T_a) \frac{dM_p}{dt} \\ + (\rho_a V_{out} C_{pa} T_{out} - \rho_a V_{in} C_{pa} T_{in}) \\ + U_c A_c (T_{am} - T_a) + [(1 - F_p)(1 - \alpha_f) \\ + (1 - \alpha_p) F_p] I_t A_c \tau_c.$$

## 2.1.3. Energy balance of the product

Rate of accumulation of thermal energy in the product = Rate of thermal energy received from air by the product due to convection + Rate of thermal energy received from cover by the product due to radiation + Rate of thermal energy lost from the product due to sensible and latent heat loss from the product + Rate of thermal energy absorbed by the product.

The energy balance on the product gives:

$$m_p(C_{pp} + C_{pl}M_p)\frac{dT_p}{dt} = A_p h_{c,p-a}(T_a - T_p) + A_p h_{r,p-c}(T_c - T_p)$$
$$+ A_p D_p \rho_p [L_p + C_{pv}(T_p - T_a)]\frac{dM_p}{dt}$$
$$+ F_p \alpha_p I_t A_c \tau_c.$$

## 2.1.4. Energy balances on the concrete floor

Rate of thermal energy flow into the floor due to conduction = Rate of solar radiation absorption on the floor + Rate of thermal energy transfer between the air and the floor due to convection. The energy balance on the concrete floor gives:

$$-k_f A_f \frac{dT_f}{dx} = (1 - F_p)\alpha_f I_t A_c \tau_c + A_f h_{c,f-a}(T_a - T_f).$$

The rate of thermal energy flow into the floor due to conduction can also be expressed as:

$$-k_f A_f \frac{dT_f}{dx} = A_f h_{d,f-g}(T_f - T_\infty).$$

## 2.1.5. Heat transfer and heat loss coefficients

Radiative heat transfer coefficient from the cover to the sky ($h_{r,c-s}$) is computed according to Duffie and Beckman (1991):

$$h_{r,c-s} = \varepsilon_c \sigma (T_c^2 + T_s^2)(T_c + T_s).$$

Radiative heat transfer coefficient between the product and the cover ($h_{r,p-c}$) is computed as (Duffie and Beckman, 1991):

$$h_{r,p-c} = \varepsilon_p \sigma (T_p^2 + T_c^2)(T_p + T_c).$$

As $h_{r,c-s}$ and $h_{r,p-c}$ are functions of temperatures, these are computed for each time interval $\Delta t$ during the simulation. The sky temperature ($T_s$) is computed as (Duffie and Beckman, 1991):

$$T_s = 0.552 T_{am}^{1.5},$$

where $T_s$ and $T_{am}$ are both in Kelvin.

Convective heat transfer coefficient from the cover to ambient due to wind ($h_w$) is computed as (Duffie and Beckman, 1991):

$$h_w = 5.7 + 3.8 V_w.$$

Convective heat transfer coefficient inside the solar greenhouse dryer for either the cover or product and floor ($h_c$) is computed from the following relationship:

$$h_{c,f-a} = h_{c,c-a} = h_{c,p-a} = h_c = \frac{Nu\, k_a}{D_h},$$

where $D_h$ is given by:

$$D_h = \frac{4WD}{2(W+D)}.$$

Nusselt number ($Nu$) is computed from the following relationship (Kays and Crawford, 1980):

$$Nu = 0.0158 Re^{0.8},$$

where $Re$ is the Reynolds number which is given by:

$$Re = \frac{D_h V_a}{v_a},$$

where $V_a$ is air speed in the dryer and $v_a$ is kinematic viscosity of air.

The overall heat loss coefficient from the greenhouse cover ($U_c$) is computed from the following relation:

$$U_c = \frac{k_c}{\delta_c}.$$

*2.1.6. Thin layer drying equation*

To obtain thin layer drying equation for copra, we used results of thin layer drying experiments of the Institute of Agricultural Engineering of Hohenheim University. The details of the dryer are described in Guarte et al. (1996). A single layer of copra was dried in this laboratory dryer under controlled condition of temperature and relative humidity. The drying experiments were conducted for temperature range of 50–70 $^oC$ and relative humidity of the drying air of 10–25% with the air speed of 0.5 m/s. The equation best fitted to experimental results is:

$$\frac{M - M_e}{M_0 - M_e} = \exp(-A_1 t^{B_1}),$$

where

$$A_1 = -0.213788 + 0.0101640T - 0.001372\text{rh}$$
$$B_1 = 1.108816 - 0.0005210T - 0.000061\text{rh},$$

where $T$ is air temperature in $^oC$ and $rh$ is relative humidity in %.

The following empirical equation developed for equilibrium moisture content ($M_e$, %db) of copra determined experimentally is used and it is given by Janjai et al. (2006a):

$$a_w = \frac{1}{1 + \left[\frac{b_0 + b_1 T}{M_e}\right]^{b_2}},$$

*2.2.8. Solution procedure*

The system of Equations are solved numerically using the finite difference technique. The time interval should be small enough for the air conditions to be constant, but for the economy of computing, a compromise between the computing time and accuracy must be considered. On the basis of the drying air temperature and relative humidity inside the drying chamber, the drying parameters $A_1$ and $B_1$ or $A_2$ and $B_2$ and the equilibrium moisture content ($M_e$) of the product are computed.
Using the $A_1$ and $B_1$ or $A_2$ and $B_2$ and $M_e$ values, the change in moisture content of the product, $\Delta M$ for all of the product for a time interval, $\Delta t$ are calculated. This system of equations is a set of implicit calculations for the time interval $\Delta t$. These are solved by the Gauss–Jordan elimination method using the recorded values for the drying air temperature and relative humidity, the change in moisture content of the product ($\Delta M$) for the given time interval from different sections of the dryer. The process is repeated until the final time is reached. The numerical solution was programmed in Compaq Visual FORTRAN version 6.5.

In the numerical calculation, the thermal and optical properties of polyethylene plate were obtained from the manufacturer. The specific heat and density of the products were experimentally determination and the properties of the air were obtained from literatures (Duffie and Beckman, 1991; ASHRAE, 1997). For other thermal properties, these were adopted from similar products reported in literatures (Mohsenin, 1980; Smitabhindu, 2008).

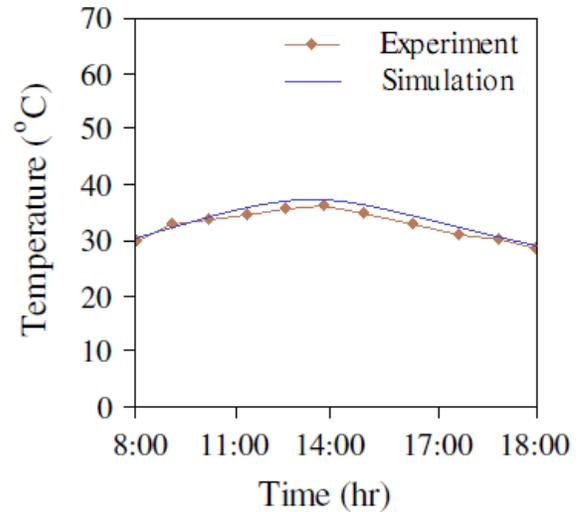

Fig. 2. Comparison of the predicted ambient temperature and experimental data.

## 3. Results and discussion

*3.1. Economic evaluation*

As there are now several units of this type of dryer are being used for production of dried copra, information used for economic evaluation is based on the field level data and recent prices of the materials used for construction of the dryers (Janjai et al., 2006b).

The price of dried products obtained from this dryer is about 20% higher than that obtained from the open-air sun drying. Approximately 250 kg of dry copra is produced annually. Based on this production scales and the capital and operating costs of the drying system of copra, the payback periods of the greenhouse solar drying system for drying of copra is estimated to be 2.3 years. This relatively short payback period is likely due to the fact that dried copra obtained from this dryer can be sold with significantly higher price than that of the products from the open-air sun drying and the dryer is used year around.

*3.2. Simulated results*

In order to validate the model, the predicted air temperatures and moisture contents of copra during drying were compared with the experimental values (Ayyappan and Mayilsamy). Fig. 3 shows a comparison between the predicted and experimental temperature values for solar drying of copra for three consecutive days. Predicted temperature shows plausible behaviour and the agreement between the predicted and observed values is reasonable. Although, there are slight discrepancies between the measured and the predicted moisture contents (Fig. 3), the difference of the prediction is only 6% and the difference for temperature prediction (Fig. 2) is 3%. This study indicates that the model can predict the temperatures with a reasonable accuracy. Furthermore, the predictions are within the acceptable limit (10%) (O'Callaghan et al., 1971).

## 4. Applications of the model

This model can be used to simulate the performance of the greenhouse solar dryer for drying of copra for different locations and climatic conditions. Also this model incorporating an economic model can be used for the optimization of the greenhouse dryer in a location of interest. The details of an optimal design of solar drying systems are given in Bala (1998).

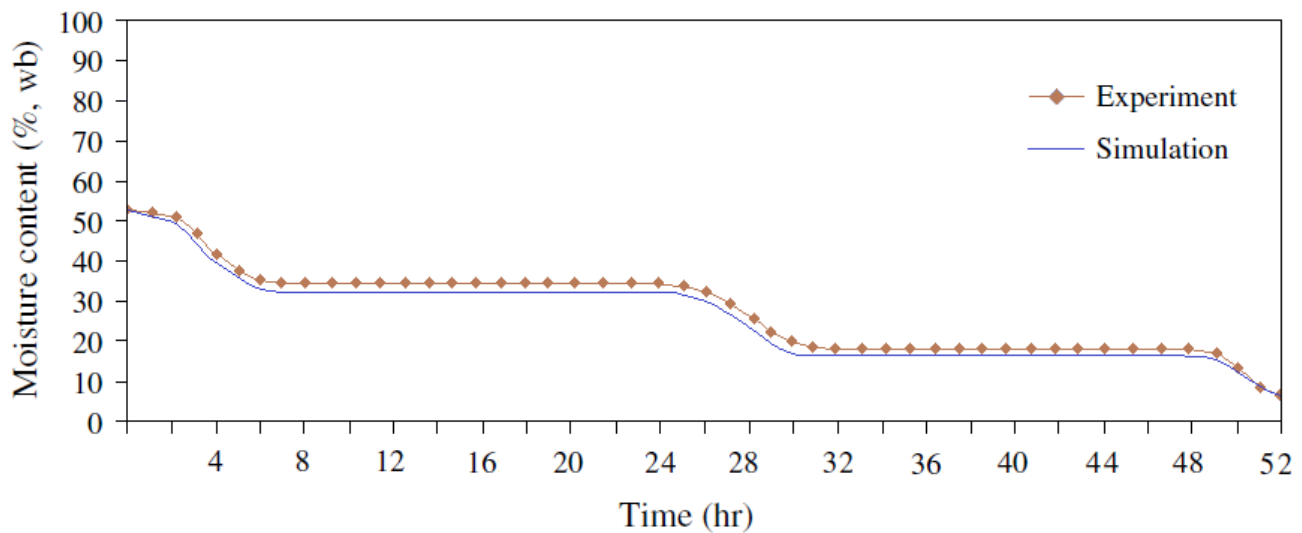

Fig. 3. Comparison of the simulated moisture content and experimental data.

## 5. Conclusions

Solar radiation varies almost sinusoidally around the peak at noon. There is significant difference in the temperatures inside the dryer with the ambient air temperatures. The patterns of changes in air velocity inside the solar greenhouse dryer follow the patterns of changes in solar radiation. Field level tests demonstrated the potential of solar drying copra in the solar greenhouse dryer. Solar drying of copra in solar greenhouse dryer resulted in considerable reductions in drying time as compared with the open-air sun drying and the products dried in the solar greenhouse dryer are high-quality dried product.

As a result of dependence on solar power, this type of dryer can be used in rural areas without electricity grids. The estimated payback periods of this greenhouse solar dryer for copra is about 2.3 years.

A system of partial differential equations for heat and moisture transfer has been developed for solar drying of copra in the solar greenhouse dryer. The simulated air temperatures inside the dryer reasonably agreed with the observed temperature data. Reasonable agreement was found between the experimental and simulated moisture contents of copra during drying and the accuracy was within the acceptable range. This model can be used for providing design data for solar greenhouse dryers and also for optimization of this type of solar dryer.

**References**

Amerine, M., Pangborn, R.M., Roessler, E.B., 1965. Principles of Sensory Evaluation of Food. Academic Press, New York, USA.

ASHRAE, 1997. Handbook: Fundamentals, American Society of Heating, Refrigerating and Air-Condition Engineers Inc. Atlanta, USA.

Bala, B.K., 1998. Solar Drying Systems. Agrotech Publishing Academy, Udaipur, India.

Ayyappan, S., Mayilsamy, K., 2010. Experimental Investigation on a Solar tunnel drier for copra drying.

Bala, B.K., Woods, J.L., 1994. Simulation of the indirect natural convection solar drying of rough rice. Solar Energy 53 (3), 259–266.

Bala, B.K., Mondol, M.R.A., Biswas, B.K., Das Choudhury, B.L., Janjai, S., 2003. Solar drying of pineapple using solar tunnel drier. Renewable Energy 28, 183–190.

Bala, B.K., Mondol, M.R.A., 2001. Experimental investigation on solar drying of fish using solar tunnel drier. Drying Technology 19, 1–10.

Barnwal, P., Tiwari, G.N., 2008. Grape drying using hybrid photovoltaic– thermal (PV/T) greenhouse dryer: an experimental study. Solar Energy 82, 1131–1144.

Condori, M., Echaz, R., Saravia, L., 2001. Solar drying of sweet pepper and garlic using the tunnel greenhouse drier. Renewable Energy 22, 447–460.

Condori, M., Saravia, L., 1998. The performance of forced convection greenhouse driers. Renewable Energy 13, 453–469.

Doiebelin, E.O., 1976. Measurement Systems. McGraw Hill, New York.

Duffie, J.A., Beckman, W.A., 1991. Solar Engineering of Thermal Processes. John Wiley and Sons, New York.

Exell, R.H.B., Kornsakoo, S., 1978. A low-cost solar rice dryer. Appropriate Technology 5, 23–24.

Esper, A., Mühlbauer, W., 1996. Solar tunnel dryer for fruits. Plant Research and Development 44, 61–80.

Hossain, M.A., Bala, B.K., 2007. Drying of hot chilli using solar tunnel drier. Solar Energy 81, 85–92.

Hossain, M.A., Woods, J.L., Bala, B.K., 2005. Optimization of solar tunnel drier for drying of chilli without color loss. Renewable Energy 30, 729–742.

Holman, J.P., 1978. Experimental Method for Engineers. McGraw Hill, New York.

Jain, D., Tiwari, G.N., 2004. Effect of greenhouse on crop drying under natural forced convection. II. Thermal modeling and experimental validation. Energy Conversion and Management 45, 2777–2793.

Jain, D., 2005. Modeling the performance of greenhouse with packed bed thermal storage on crop drying application. Journal of Food Engineering 71, 170–178.

Janjai, S., Hirunlabh, J., 1993. Experimental study of a solar fruit dryer. In: Proceedings of ISES Solar World Congress, vol. 8, Biomass, Agriculture, Wind, pp. 123–128.